\documentclass[11pt]{IEEEtran}

\usepackage{graphicx}
\usepackage{fancyhdr}
\usepackage{xcolor}
\usepackage{hyperref}

\begin{document}

\bibliographystyle{IEEEtran}


\newcommand{\exsys}[1]{
  \vskip 2 ex 
  \noindent
  \textbf{#1.~}
}

\newcommand{\figureSingle}[1]{
  \begin{figure}[t]{\sloppy \footnotesize#1}\end{figure}
}

\newcommand{\figurePair}[2]{
  \begin{figure}[t]
    \begin{minipage}[t]{.48\columnwidth}{\sloppy \footnotesize#1}\end{minipage}
    \hfill
    \begin{minipage}[t]{.48\columnwidth}{\sloppy \footnotesize#2}\end{minipage}
  \end{figure}
}

\newcommand{\tableSingle}[1]{
  \begin{table*}[t]{\sloppy \footnotesize#1}\end{table*}
}

\newcommand{\frc}{\scriptsize N}
\newcommand{\acc}{\scriptsize m$/$s\tiny$^2$}
\newcommand{\mss}{\scriptsize kg}
\newcommand{\spU}{\rule{0in}{3ex}}
\newcommand{\spD}{\rule[-1ex]{0in}{1ex}}
\newcommand{\spB}{\rule[-1ex]{0in}{4ex}}


\newcommand{\pad}{
  \hskip .3in
}

\title{Combining Active and Passive Simulations for Secondary~Motion}
\author{James F. O'Brien \pad Victor B. Zordan \pad Jessica K. Hodgins \\
\normalsize{~}\\
\normalsize{College of Computing and Graphics, Visualization, and Usability Center}\\
\normalsize{Georgia Institute of Technology}\\
\normalsize{Atlanta, GA 30332-0280}\\
\normalsize{{\small \tt $[$obrienj$|$victor$|$jkh$]$@cc.gatech.edu}}\\
}

\maketitle
\thispagestyle{empty}


\begin{abstract} 
  Varied, realistic motion in a complex environment can bring an
  animated scene to life.  While much of the required motion comes
  from the characters, an important contribution also comes from the
  passive motion of other objects in the scene.  We use the term
  \textit{secondary motion} to refer to passive motions that are
  generated in response to environmental forces or the movements of
  characters and other objects.  For example, the movement of clothing
  and hair adds visual complexity to an animated scene of a jogging
  figure.  In this paper, we describe how secondary motion can be
  generated by coupling physically based simulations of passive
  systems to active simulations of the main characters.  

  We categorize interactions according to how forces are communicated
  between the objects.  In a \textit{two-way coupled} system both
  objects experience forces generated by the interaction.  In a
  \textit{one-way coupled} system, one object is designated as
  \textit{primary} and the other \textit{secondary}.  The interaction
  forces are only applied to the \textit{secondary} object allowing
  the primary object to be simulated independent of the secondary.
  Between these two solutions are a variety of \textit{hybrid} solutions 
  where the interaction model is approximate.  This range of options 
  allows the animator to make an appropriate tradeoff between accuracy
  and computational speed.  To demonstrate when each kind of coupling is most
  appropriate, we present several examples including a gymnast on a
  trampoline, a man on a bungee cord, a stunt kite, a gymnast landing
  on a flexible mat, a diver entering the water, and several human
  figures wearing clothing.  The information gained from analyzing
  these examples is summarized in a set of guidelines for coupling
  active and passive systems.
  
  \textit{\color{red} Author's reprint. Published as James F. O'Brien, Victor B. Zordan, and Jessica K. Hodgins. ``Combining Active 
  and Passive Simulations for Secondary Motion''. IEEE Computer Graphics and Applications, 20(4):86--96, 2000. }
\end{abstract}

\pagestyle{fancy}
\fancyfoot{}
\fancyhead{}
\fancyfoot[RO,LE]{\thepage}
\fancyhead[LE]{\fontsize{8}{8}\textsf{Combining Active 
  and Passive Simulations for Secondary Motion}}
\fancyhead[RO]{\fontsize{8}{8}\textsf{IEEE Computer Graphics and Applications, 20(4):86--96, 2000.}}




\section{Introduction}

Objects that move in response to the actions of a main character often
make an important contribution to the visual richness of an animated
scene.  We use the term \textit{secondary motion} to refer to passive
motions that are generated in response to the movements of characters
and other objects or environmental forces.  Secondary motion may be
created by background elements or by objects interacting with an
active character.  The flags shown in Figure~\ref{figureFlags} are
examples of secondary motion generated by environmental forces, while
the trampoline and skirt in Figure~\ref{figurePlayGroundImage} are
objects that exhibit secondary motion in response to the actions of
active characters.

Secondary motions are not normally the main focus of an animated
scene, yet their absence can distract or disturb the viewer,
destroying the illusion created by the scene.  For example, if the
skirt in Figure~\ref{figurePlayGroundImage} were rigid, the scene
would be less believable; with painted-on, skin-tight clothing, the
scene would be less interesting.  While the viewer may not always be
explicitly aware of secondary motions, they are an important part of 
many animated scenes.

\newcommand{\figureFlags}{
  \centerline{\includegraphics[width=\columnwidth]{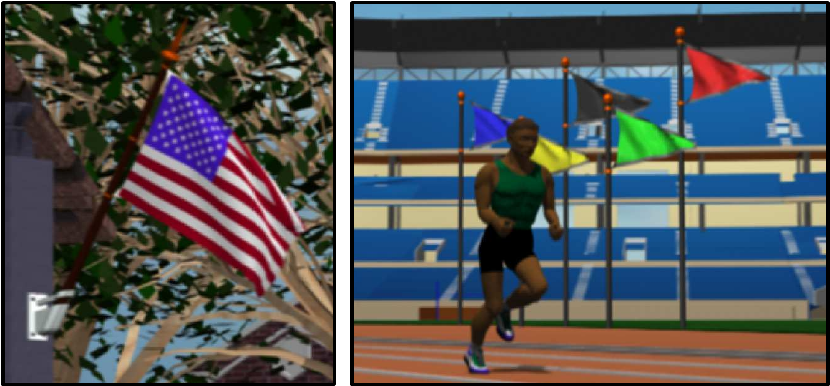}}
  \caption{
    \textbf{Simulated flags in the wind. } Flags are examples of simple
    background elements that move in response to environmental effects
    such as wind.
  }\label{figureFlags}
}
\figureSingle{\figureFlags}

\newcommand{\figurePlayGroundImage}{
  \typeout{                                                                        }
  \typeout{------ Don't forget to replace figurePlayGroundImage w/ high res. ------}
  \typeout{                                                                        }
  \centerline{\includegraphics[width=\columnwidth]{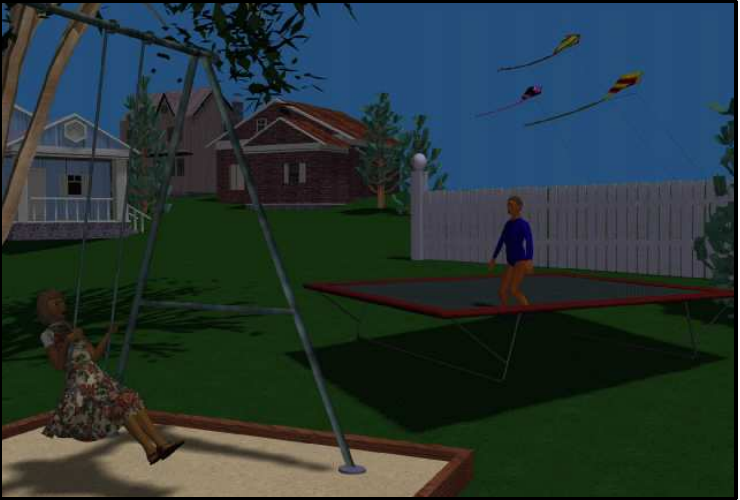}}
  \caption{
    \textbf{An animated scene with secondary motion.}  Both the swinger's
    skirt and the bed of the trampoline must move if the animation is to
    be convincing.  Additional moving elements, such as the kites flying in
    the wind, further enhance the realism of the scene.
  }\label{figurePlayGroundImage}
}
\figureSingle{\figurePlayGroundImage}

Much of the research in computer animation has focused on the
difficult problem of animating the primary characters.  Because
objects that exhibit secondary motions tend to be complex, deformable
objects with many degrees of freedom, the techniques that have been
developed for character animation are usually not appropriate for
animating secondary motion.  In particular, methods based on motion
capture or key framing are often impractical for secondary motion. As
a result, specialized procedural methods have been developed for many
of these objects.

While procedural models may be derived in a number of ways, physically
based simulation has proven to be both a highly effective and an
elegant solution, particularly for passive systems with many degrees
of freedom.  One advantage of simulation is that the motion is
generated automatically from the initial specification of the
environment and the applicable physical laws.  For some applications,
such as character animation, this automation results in an undesirable
loss of direct control over the details of the motion.  However, for
secondary motion this lack of control is usually not a significant
problem because these motions are passive, dictated only by forces
from the environment or the actions of the primary characters.  Even
in situations where aesthetic considerations call for an exaggerated
or otherwise unrealistic motion, it is most often the movement of the
actor that is exaggerated while the passive secondary motions simply
respond to the exaggerated motion.

Simulation has been successfully used to model many isolated
phenomena, but secondary motion by definition involves interactions
between objects.  Specialized simulations can be coupled together
using inter-system constraints and forces to model the complex
interactions that occur in the real world.  The primary contribution
of this work is an exploration of the issues involved when passive
secondary systems are coupled to another, primary, system. Typically,
but not necessarily, the primary system will be active, having an
internal source of energy and a control system to govern its behavior.

We classify methods for coupling two systems together as
\textit{two-way}, \textit{one-way}, or \textit{hybrid}.  To clarify
the differences between these three forms of coupling, we use the
interaction between a basketball (primary) and net (secondary) as an
illustrative example.  If the simulations are two-way coupled, the
rotational and linear velocity of the ball will be changed by the
contact with the net, and the net will be pushed out of the way by the
ball.  If the coupling is one-way, the motion of the ball is not
affected by the net, and the ball continues on a ballistic trajectory.
The deformation of the net will be more extreme than in the two-way
coupled case and the motion will not match that of an actual
basketball and net as closely.  Between these two solutions are a
variety of hybrid solutions where the interaction model is
approximate.

The physics of a particular situation and the fidelity of the required
motion determine how the simulations should be coupled.  In some
situations, one-way or hybrid coupling can result in substantial
computational savings with little loss of realism; in others, a tight
two-way coupling is essential.  To illustrate some of these issues,
and to demonstrate the generality of our approach for generating
secondary motion, we consider several systems that are built by
coupling simulated components: a gymnast on a trampoline, a man on a
bungee cord, a flying stunt kite, a gymnast landing on a flexible mat,
a diver entering the water, and several human figures wearing
clothing.


\section{Background} \label{BackgroundSection}

A number of techniques have been developed that use physically based
simulation to generate motion for animation.  Most of the research has
focused on the issue of designing a simulation method for a particular
type of phenomenon or motion and, with the exception of work by Baraff
and Witkin\,\cite{Baraff:1997:PD}, techniques for coupling simulations
have been largely unexplored.  This section discusses techniques for
simulating passive and active systems as well as previous work related
to combining systems.

Simulation has proven particularly successful in animating passive
systems with many degrees of freedom such as cloth, water, hair, and
other natural phenomenon.  Cloth simulation packages are even
appearing in commercial packages and clothing simulation was used
successfully in the Oscar winning short Geri's
Game\,\cite{DeRose:1998:SSC}.  Many of the techniques developed to
model cloth are based on the spring and mass techniques originally
introduced to the animation community by Terzopoulos and his
colleagues in 1987\,\cite{Terzopoulos:1987:EDM}.  Other cloth systems
have been developed since that are based on finite element methods and
include self-collision as well as interaction with synthetic actors
(for example~\cite{Volino:1995:VET}).  Recent work has shown
interactive cloth simulation to be feasible\,\cite{Baraff:1998:LSC}
and simple cloth objects are appearing as effects in physically based
electronic games.

Most of the water models presented in the literature focus on such
specific phenomenon as splashing, waterfalls, and spray.  The
techniques provide varying levels of realism and interaction with
external objects.  Highly realistic results have been achieved using a
variation of 3D Navier-Stokes equations to animate liquids in complex
environments\,\cite{Foster:1996:RAL}.  As with cloth simulation, fluid
simulation techniques have progressed to the point where both
interactive simulations\,\cite{Stam:1999:SF} and their use in
commercial productions\,\cite{Witting:1999:CFD} are feasible.

Other natural phenomenon have been modeled including wind and
atmospheric effects, deformable terrain, and natural hair motion.
Some of these systems are combined with external elements to generate
secondary motion.  Li and Moshell modeled soil slippage and
manipulation\,\cite{Li:1993:MSR}.  Their system supported interaction
through a controllable bulldozer and other earth--moving equipment.
Sumner, O'Brien, and Hodgins introduced a system for animating
deformable terrain to create imprints from simulated characters in
sand, snow, and mud\,\cite{Sumner:1999:ASM}.  Physical models has also
been used to model how objects fracture.  O'Brien and Hodgins have
developed a finite element technique for modeling deformable objects
that can break, crack, or tear when they deform in response to
external forces\,\cite{OBrien:1999:GMA}.

The use of simulation for active systems is not as widespread as it is
for passive systems because robust control algorithms that produce
natural-looking motions are difficult to design with existing
techniques.  A number of hand-tuned simulations for rigid-body human
and deformable non-human characters have been introduced
(see, for example,\,\cite{Badler:1993:SHC,Hodgins:1995:AHA,Tu:1994:AFP}).

Some of the work on passive systems includes specific examples of
coupling two systems together.  For example, combining deformable
clothing with the motions of synthetic
actors\,\cite{Carignan:1992:DAS} and manipulating soil with a
bulldozer\,\cite{Li:1993:MSR} are similar to what we term one-way
coupling.  However, the general concept of coupling was not
investigated in these papers, and responsive active simulations, as
would be the case for two-way coupling, were not considered.

The work of Baraff and Witkin\,\cite{Baraff:1997:PD} is most closely
related to the work presented in this paper.  They present a method
for combining groups of passive systems including particle, clothing,
and passive rigid-body models.  Their work focuses on a method that
uses constraints to allow multiple systems to interact.  They include
examples of complex interaction such as two-way coupling between a
stack of rigid objects and a cloth object or particle spray.  In our
work, we focus on higher-level issues including when coupling two
systems is appropriate, how approximations can be introduced to
increase interactivity and efficiency without significantly degrading
the results, and issues specific to coupling active systems to passive
ones.


\section{Coupling} \label{CouplingSection}

Our goal is to combine simulations of individual objects or phenomena
so that they can interact with each other to produce secondary motion.
The techniques referenced in Section~\ref{BackgroundSection} address
modeling the behavior of particular objects or phenomena using
specific simulation techniques, and we build on this existing work.
Thus, we adopt a modular approach where two or more systems are
coupled together and emphasize the design of the interfaces between
these systems.

Forces applied between the systems provide a natural way for one
simulation to interact with another.  We group the interactions into
three categories based on the method of approximating the inter-system
forces: two-way coupled, one-way coupled, and hybrid.

In the remainder of this section, we illustrate the differences
between these coupling techniques, with an example of a basketball
going through a net.  In this simple example, the primary system is
the basketball and the net is the secondary system.  The collisions
between the net and the ball are the interactions that we aim to
model. The ball is modeled as a spherical rigid body that is free to
translate and rotate in space.  The ball is initialized with a linear
and angular velocity determined by the animator and once in flight it
experiences a gravitational acceleration.  The net is modeled with a
spring and mass network that is attached by springs to a fixed hoop
rim.  The mass points experience forces due to gravity, the actions of
the springs, and their interaction with the ball.


\subsection{Two-Way Coupled} \label{ssec-twoway}

While any computer simulation involves some level of approximation, a
two-way coupled simulation is designed to model the interaction as
realistically as possible given the component systems.  Two-way
interactions affect both components, and the forces applied to one
system are mirrored by equal and opposite forces applied to the other.
The systems are simulated in lock step with each other and the actions
of each system directly affect the other.

We implemented the basketball and net as a two-way coupled system.  To
model the interaction forces, collision constraints are imposed to
prevent the points of the net's mesh from penetrating the surface of
the basketball.  When contact is detected, a constraint force prevents
further penetration, a stabilizing damping force absorbs a portion of
the impact energy, and a restoring force corrects any penetration
error.  Friction is implemented using a Coulomb friction model.  The
resultant force is applied to the appropriate points of the net and an
equal and opposite force, with a corresponding moment, is applied to
the ball.

\newcommand{\figureSimBalls}{
  \centerline{\includegraphics[width=\columnwidth]{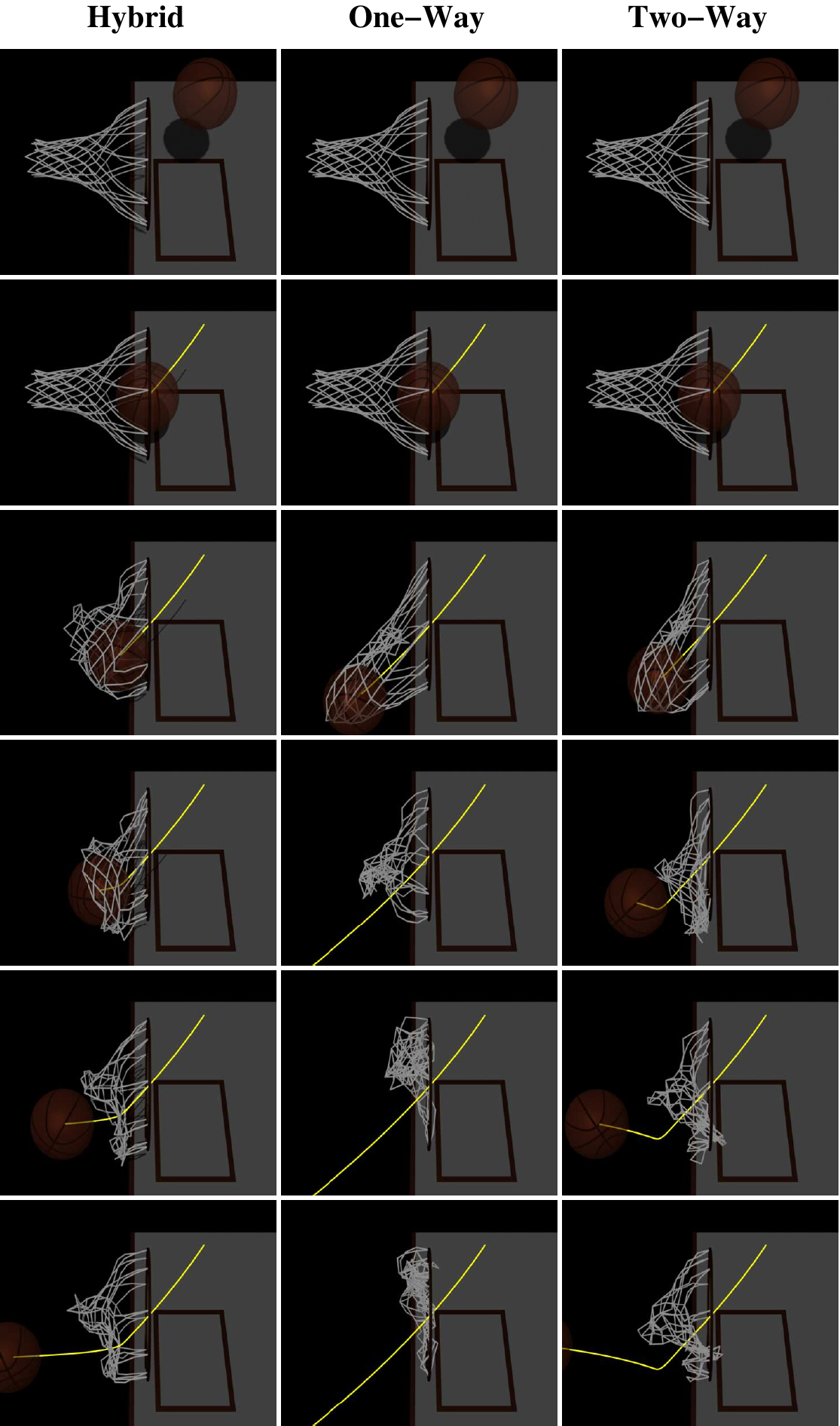}}
  \caption{
     \textbf{Simulated basketball and net with different couplings.}
     The yellow line highlights the path of the ball generated using
     two-way, one-way, and hybrid coupling. Images are sampled at
     $0.0833$\,second intervals.
  }\label{figureSimBalls}
}
\figureSingle{\figureSimBalls}

The path of the ball with two-way coupling is shown in the top
sequence of Figure~\ref{figureSimBalls}.  The ball enters the net at a
shallow angle while spinning clockwise, causing the net to deflect
from its rest configuration until the strings in the net are pulled
tight and the sideways velocity of the ball is reduced.  The ball
drops out of the net with a substantially altered trajectory.

The main drawback to this type of coupling is the computation time
required before the path of the ball may be viewed.  The parameters
for the net have been chosen to represent nylon cord which, while
light and flexible, is highly resistant to stretching.  As a result,
the simulation of the net is numerically stiff and requires a small
time step, on the order of $10^{-5}$\,seconds.  Additionally, the net
contains hundreds of masses, each of which must be integrated for
every time step.  The ball, on the other hand, is a rigid body with
isotropic inertial moments and its ballistic flight may be simulated
with arbitrarily large time steps.  Therefore, computing a time step
for the ball is computationally much less expensive than computing a
time step for the mesh.  If simulated alone, the ball could be
computed in real time on even very modest machines, allowing the
animator to interactively view and refine the motion by changing the
initial conditions.  When the two simulations are coupled together,
the animator must wait several minutes before the motion can be
viewed.  We refer to the time between specifying parameters and
viewing the results as the \textit{debug cycle time}.

Finally, the total computation time required to calculate the result
of the two-way coupled simulation may be greater than the total time
required to simulate each of the systems separately, even allowing for
the additional work required to compute the interaction.  For example,
consider coupling two simulations where one system has a small
computational cost per time step but requires small time steps, and
where the other system has a large cost for each time step but is
stable at large time steps.  Either system alone may be fast enough to
be usable, but when the two systems are combined, the poor stability
of the first is likely to dictate a small time step for both thus
greatly increasing the total computational cost.


\subsection{One-Way Coupled}

With a one-way coupled system, the interaction forces are applied only
to the secondary system, leaving the primary system unaffected by the
interaction.  This approach relies on the assumption that the
neglected forces would have a minimal effect on the primary system if
they were applied.  This situation is likely to occur when the mass of
one component system is several times the mass of the other, when one
system is constrained in a way that would counteract the interaction
forces, or when an active primary system would be able to trivially
correct for any disturbances caused by the interaction with the
secondary system.

We implemented the basketball and net as a one-way coupled system to
illustrate this coupling technique.  The interaction forces are
computed as before, but no forces are applied to the ball.  The
resulting motion can be seen in the center row of
Figure~\ref{figureSimBalls}.  The ball's path is not affected by the
net and the resulting trajectory is ballistic and therefore
unrealistic.  The net is forced to stretch a great deal, despite its
stiff material parameters, eventually causing a violation of the
collision constraints (fifth image of second row in
Figure~\ref{figureSimBalls}).  Because the net is substantially
deformed by the interaction, it requires a significantly smaller time
step to prevent instability.

One benefit derived from this method of coupling is that the two
systems may be simulated separately, potentially avoiding a long
primary debug cycle time.  In general, this type of coupling is easier
to implement than two-way coupling because only the secondary system
is modified.  It also allows coupling where it is not possible or
desirable to modify the primary system such as in the case of a motion
capture-driven or hand-animated character.

If the assumption that the interaction would have had a minimal effect
on the primary system is wrong, as in the basketball and net example,
the resulting motion will appear unrealistic.  Even in cases where the
effect would have been quite subtle, the resulting motion can appear
incorrect in a way that most viewers might not notice consciously, but
may nonetheless find disturbing and distracting.


\subsection{Hybrid}

A hybrid system is a compromise between the accuracy of two-way
coupling and the speed of one-way coupling.  As with one-way coupling,
the primary system is computed independently of the secondary system.
However, rather than ignoring the effect of the interaction on the
primary system entirely, a simple approximation of the secondary
system, a stand-in, interacts with the primary system.  The motion of
the primary system is then used to drive the secondary system as in
the one-way coupled case.

The bottom row of Figure~\ref{figureSimBalls} shows the results of
implementing the basketball and net with a hybrid coupling.  The
effect of the net on the ball is approximated with a damping field
co-located with the rest configuration of the net.  As the ball passes
through the field, its translational and rotational momentum are
damped according to parameters selected by the animator.  Once the
path of the ball has been determined, the net is simulated using the
generated ball path.

The motion of the ball generated with the hybrid system is
substantially different from the motions generated by the two- and
one-way coupled simulations.  The ball's horizontal and rotational
velocity are slowed significantly, but, in contrast to the velocities
seen with two-way coupling, they do not reverse because the simple
approximation of a damping field is not capable of producing that
behavior.  The ball's path and the motion of the net are, however,
qualitatively similar to that seen with two-way coupling.  The path
generated by the hybrid simulation is also significantly different
from the parabolic trajectory of the one-way coupled system and the
net is not stretched in an unrealistic fashion.

This example uses a relatively simple stand-in to model the action of
the net on the ball, but an arbitrarily realistic model could be used
for the stand-in.  The distinction is that while the simulation of the
secondary system must model all the visible behaviors of the secondary
object, the stand-in need only approximate the desired interactions.
Designing a stand-in that can be simulated quickly and efficiently is
much easier than designing a secondary system that can be fully
coupled to the primary system.  For any given secondary system, there
are many possible stand-ins with various levels of physical realism,
and the appropriate stand-in depends on the level of realism required
by the interaction.

The design and parameters of the approximation used for the hybrid
simulation provide additional control handles for the animator.  For
the basketball example, the location, size, and damping constants of
the field can be adjusted to achieve a desirable path for the ball.
Because hybrid coupling should provide a shorter debug cycle time for
the primary system than two-way coupling, the animator may
interactively adjust these parameters until the desired trajectory is
achieved.


\subsection{Comparison: Simulated versus Real World}

In the above discussion, we referred to the two-way coupled simulation
as the most realistic of the three methods and implicitly used it as a
standard against which the results of the other two methods were
compared.  The true standard, however, is the motion of a real
basketball and net.  Figure~\ref{figureCompLiveSim} compares images
from video footage to rendered images of the two-way coupled
simulation with similar initial conditions.  The simulated ball and
net move in a way that closely resembles the motion shown in the video
images.

\newcommand{\figureCompLiveSim}{
  \centerline{\includegraphics[width=\columnwidth]{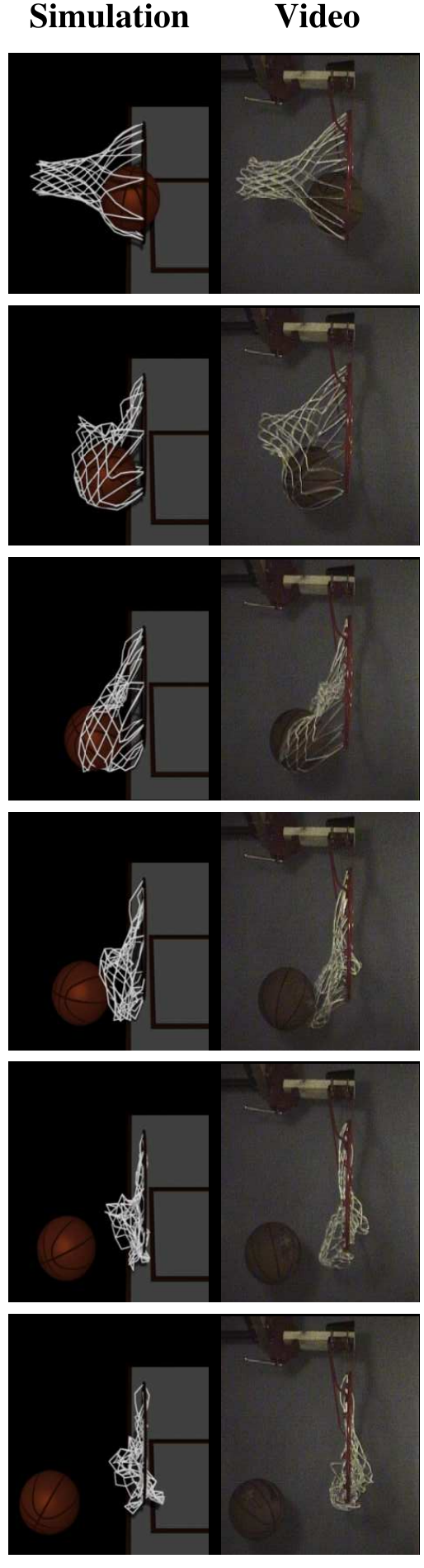}}
  \caption{
    \textbf{Comparison between simulation results for two-way coupling and
    video footage.}  The top row of images shows frames captured from
    video footage of a real basketball and net.  The bottom row shows
    a two-way coupled simulation with matching initial conditions.
    Images are sampled at $0.067$\,second intervals.
  }\label{figureCompLiveSim}
}
\figureSingle{\figureCompLiveSim} 


\section{Example Systems} \label{ExampleSection}

In this section, we discuss a variety of examples and how they can be
implemented as coupled systems.  In the previous section, we chose to
use the basketball and net system as an illustrative example because
it is a familiar system that is relatively simple to work with and we
were able to implement it using each of the three coupling methods.
For each of the examples in this section, we discuss a single
implementation that employs the most suitable coupling technique.  The
examples are presented in approximately ascending order of complexity.
We focus on passive systems that are modeled with mass and spring
systems or with simplified fluid dynamics models, however, the ideas
we describe should be applicable to other types of physically based
systems.


\exsys{Leaves} 
Figure~\ref{figureLeaves} shows leaves blowing in the wind.  The
bicyclist generates a wind field that stirs up leaves in the road as
he moves past them.  We use Wejchert and Haumann's simplified
aerodynamics model to drive the motion of flexible leaves blowing in
the wind\,\cite{Wejchert:1991:AA}.  Because the actor does not
experience any forces due to the motion of the leaves, the system is
one-way coupled.

\newcommand{\figureLeaves}{
  \centerline{\includegraphics[width=\columnwidth]{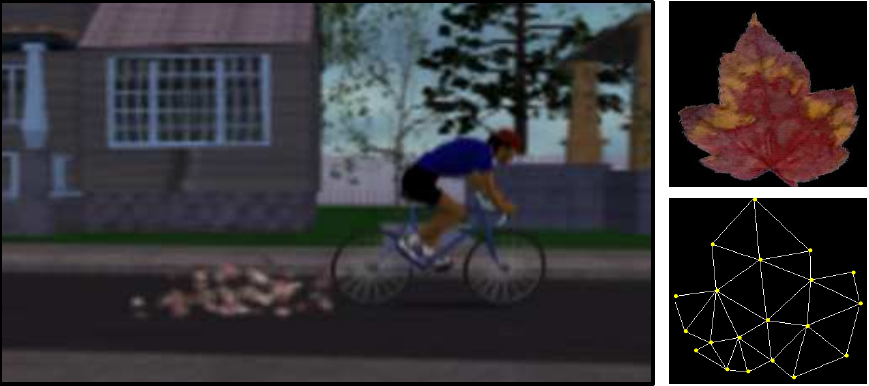}}
  \caption{
    \textbf{Spring and mass model of a leaf. }  The leaves are
    influenced by wind fields generated by moving objects such as a
    bicyclist.  The diagram on the right shows the texture map and the
    spring and mass network used to model a leaf.  For clarity,
    additional springs that resist bending and shear are not shown.
  }\label{figureLeaves}
}
\figureSingle{\figureLeaves}


\exsys{Clothing}
We have modeled clothing as a one-way coupled system.  This choice is
appropriate because the effect of the clothing on the simulated human
is negligible.  The clothing is modeled with a mass and spring system
that is generated automatically from a geometric model.  Collisions
between the clothing and the actor are detected by intersecting the
triangle faces of the actor's polygonal model with the triangles of
the clothing model.  Figure~\ref{figureWearClothes} shows a runner
wearing a tee-shirt and sweat pants and a child on a swing wearing a
skirt.

\newcommand{\figureWearClothes}{
  \centerline{
    \hfill ~ 
    \hfill
    {\includegraphics[height=50mm]{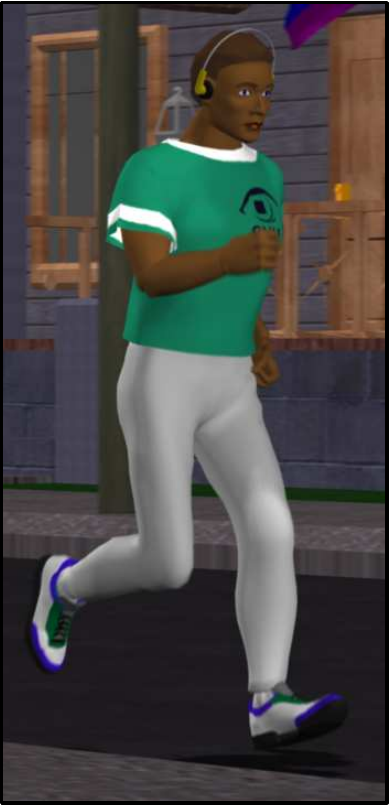}}
    \hfill
    {\includegraphics[height=50mm]{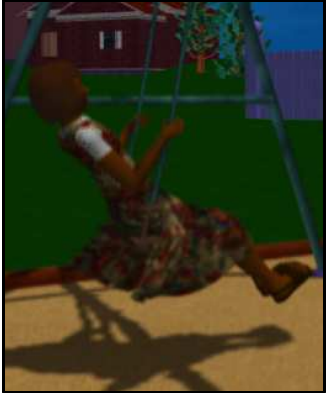}}
    \hfill ~ 
    \hfill
  }
  \caption{
    \textbf{Synthetic actors wearing simulated clothing.}
    While the clothing worn by the actors moves in response to their
    actions, the effect of the clothing on the runner and child on the
    swing is assumed to be negligible.  
  }\label{figureWearClothes} 
}
\figureSingle{\figureWearClothes}


\exsys{Floor Mat}
Figure~\ref{figureLandOnMat} shows a gymnast landing on a deformable
floor mat after performing a handspring vault.  The floor mat makes
the scene appear more realistic by softening the landing and by
deforming to create a visual connection between the gymnast and the
rest of the scene.

\newcommand{\figureLandOnMat}{
  \centerline{\includegraphics[width=\columnwidth]{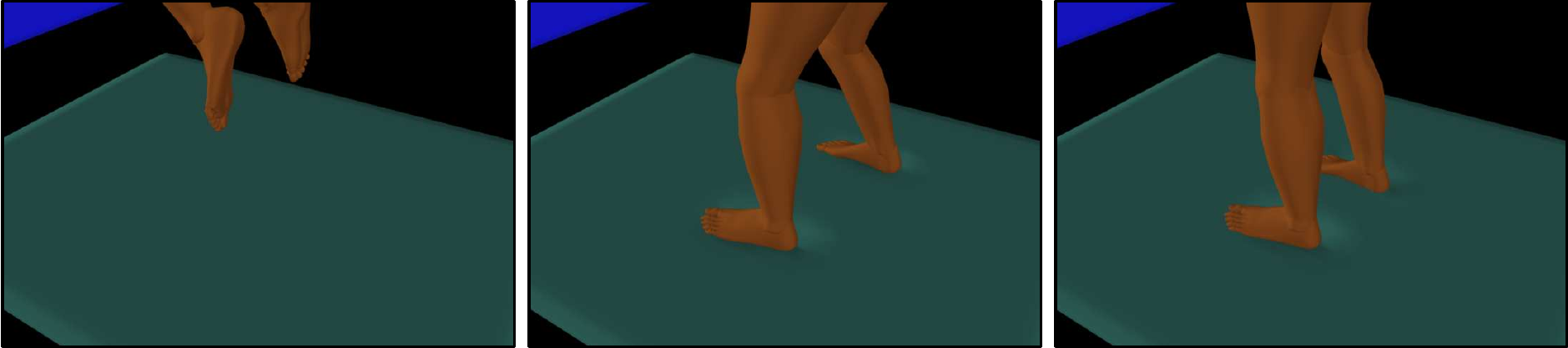}}
  \caption{
    \textbf{Gymnast landing on a deformable mat. } This closeup shows 
    a gymnast landing on a deformable floor mat after a hand-spring
    vault.  The give of the mat prevents the landing from having a
    painful, bone jarring appearance and the subsequent deformation
    creates an important connection between the actor and the
    background.
  }\label{figureLandOnMat}
}
\figureSingle{\figureLandOnMat}

The floor mat is modeled using a mass and spring system and the
gymnast is modeled with a hierarchy of rigid bodies governed by an
active control system.  Because the gymnast's controller is tuned by
hand, a quick debug cycle time is important.  The gymnast simulation
is relatively fast and can be run interactively, but the mat
simulation is several times slower.  Using a two-way coupling to link
these systems would result in an unacceptably slow debug cycle time
for the gymnast, but a one-way coupling would not have the desired
result of softening the landing.  Instead, we use a hybrid solution.
The forces applied to the gymnast's feet are computed as if she were
landing on a grid of vertical springs.  Although this simple model
will not capture subtle effects, such as sideways slip, the
approximation has the desired result of softening the landing while
still being very fast to simulate.  Once the gymnast's motion has been
computed, it is used to drive the floor mat simulation and produce the
desired deformation of the mat.


\exsys{Water} 
We have used one-way, two-way, and hybrid couplings to combine rigid
body models with a height field based water simulation
technique\,\cite{OBrien:1995:DSS}.  Figure~\ref{figureFootSplash}
shows a runner stepping in a puddle.  Because his motion is not
significantly affected by the water, a one-way coupling is used to
model the interaction.  On the other hand, a diver entering the water
from a $10$\,meter platform (Figure~\ref{figureDiverSplash}) should be
significantly affected by the water although the degree to which the
viewer is able to observe this effect is limited.  Therefore we use a
hybrid coupling where the diver encounters a viscous damping field
that exerts drag forces on the body parts that are under water.  The
resulting motion is then used to drive the water simulation.  Objects
floating on the surface of a pond, on the other hand, require two-way
coupling because the water's motion is affected by the motion of the
floating objects and their motion is in turn affected by the water
(Figure~\ref{figureBallDrop}).

\newcommand{\figureFootSplash}{ 
  \centerline{\includegraphics[width=\columnwidth]{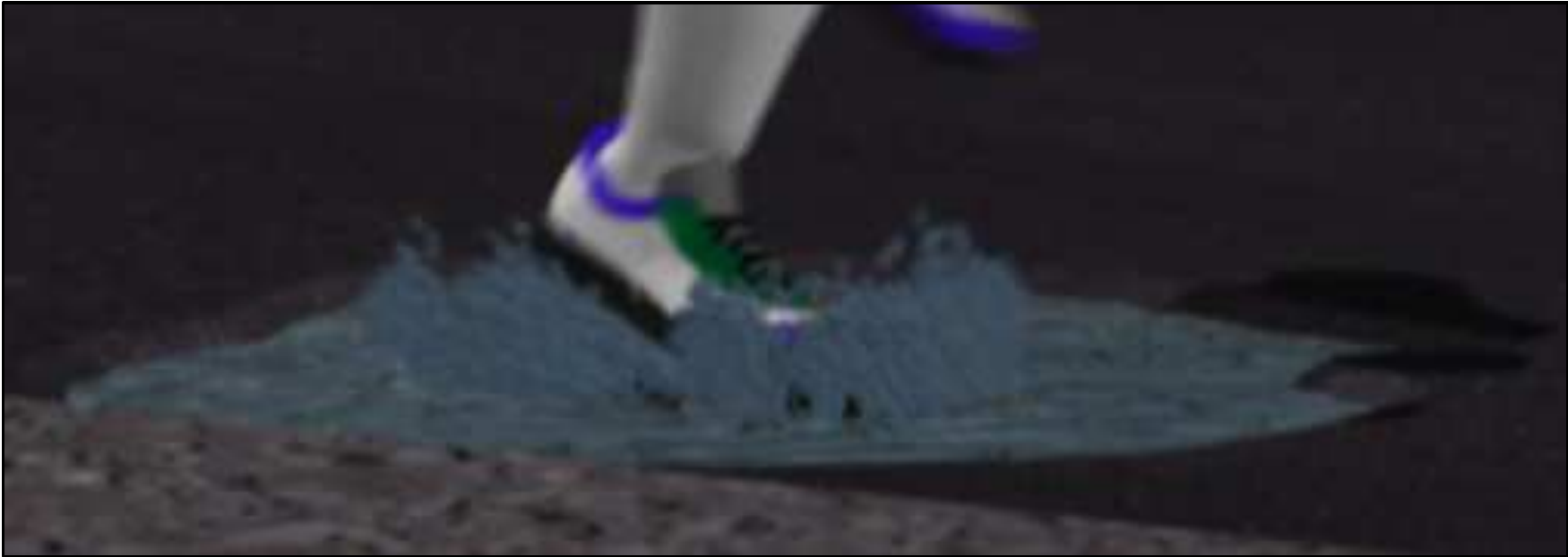}}
  \caption{
    \textbf{Foot stepping in a puddle of water.}  Although the runner's
    motion is unaffected, the impact of the foot causes a splash.
  }\label{figureFootSplash}
}
\figureSingle{\figureFootSplash}

\newcommand{\figureDiverSplash}{
  \centerline{\includegraphics[width=0.66 \columnwidth]{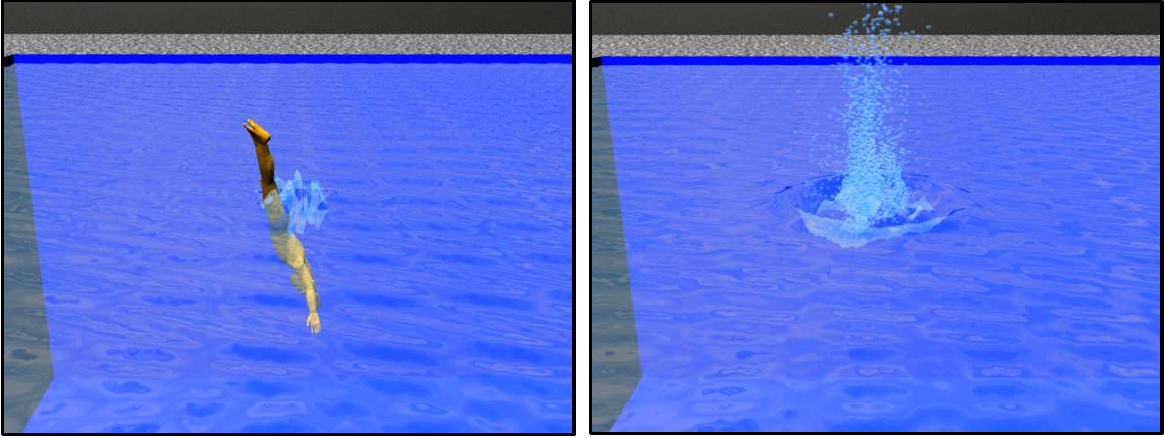}}
  \caption{
    \textbf{Diver entering the water.} As the diver enters the water, he
    slows down due to viscous drag and creates a splash. 
  }\label{figureDiverSplash}
}
\figureSingle{\figureDiverSplash}

\newcommand{\figureBallDrop}{
  \centerline{\includegraphics[width=\columnwidth]{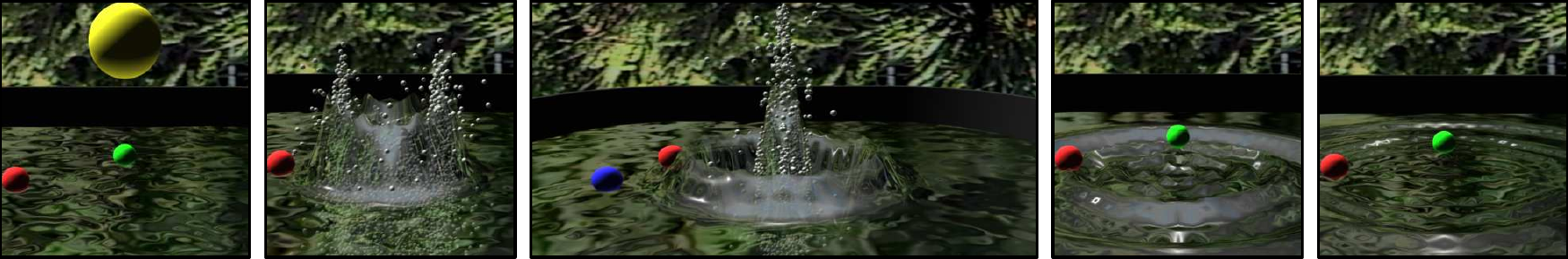}}
  \caption{
    \textbf{Balls floating in water.}  Two-way coupling is used to
    model the interaction between floating balls and water in a small
    pond.  When the lighter balls are dropped into the water, they
    create small disturbances and float on the surface.  The larger,
    more dense ball creates a larger disturbance and sinks.  The
    motions of the floating balls are affected by the resulting waves.
  }\label{figureBallDrop}
}
\figureSingle{\figureBallDrop}


\exsys{Kites and Stunt Kite} In addition to modeling the interactions
between separate objects, two-way coupling can also be used to model
the interactions of different components within a single object.  By
separating the object into components, we can make simplifications
that are consistent with the specific qualities desired in the
resulting motion of each component.  We have used this approach to
model single and double line kites flying in the air.  We divide each
kite into four components: cloth wing, frame, bridle and string, and
tail, as shown in Figures~\ref{figureKitesDiag}
and~\ref{figureKitesInSky}.

\newcommand{\figureKitesDiag}{ 
  \centerline{\includegraphics[width=\columnwidth]{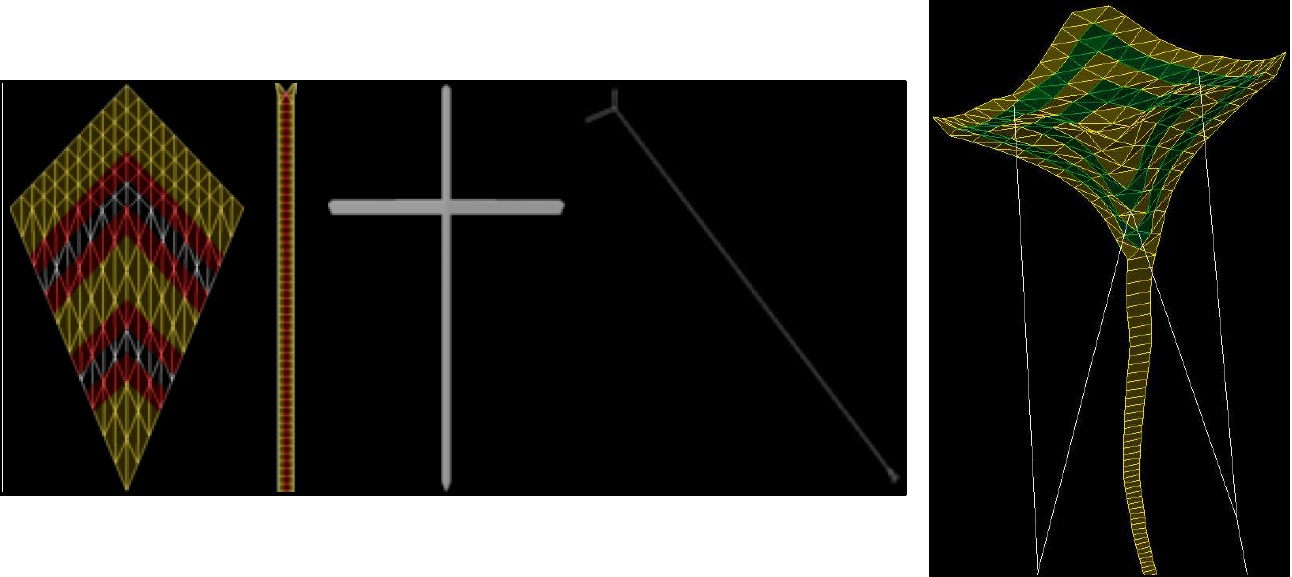}}
  \caption{
    \textbf{Diagram of kite assembly.}  The four figures on the left show
    the components of the single line kite: the wing, tail, frame, and
    line.  On the right, the two-line stunt kite is shown assembled.
  }\label{figureKitesDiag}
}
\newcommand{\figureKitesInSky}{
  \centerline{\includegraphics[width=\columnwidth]{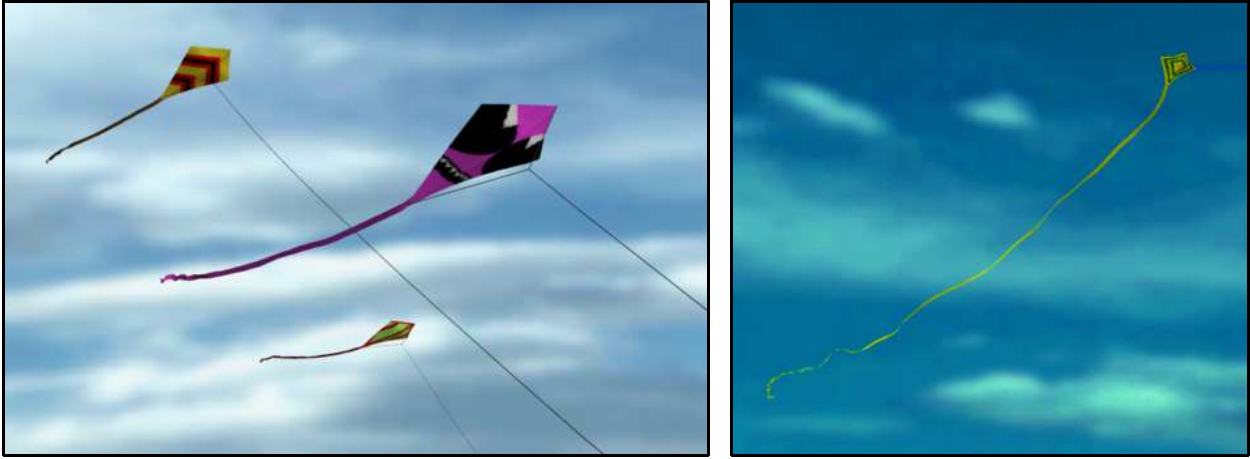}}
  \caption{
    \textbf{Kites in the sky. } The image on the left shows
    three single line kites in the air.  The image on the right shows
    the two-line stunt kite as it performs a looping maneuver.
  }\label{figureKitesInSky}
}
\figurePair{\figureKitesDiag}{\figureKitesInSky}

The kite is held aloft in the presence of gravity by the combined
action of a horizontal wind field and the tension in the string.  Lift
and drag forces are generated on the wing and tail using the same
simplified aerodynamic model as was used for the leaves.  The wing
ripples and deforms as the wind acts on it, causing variations in the
net aerodynamic forces that propagate to the frame and creating subtle
variations in the kite's motion.  The drag on the tail serves to
stabilize the system.  The lower ends of the strings on the double
line stunt kite are moved by a control system that directs the path of
the kite much as a person would fly a real stunt kite.


\exsys{Bungee Jumper} The bungee jumper shown in
Figure~\ref{figureBungee} is an example of a two-way coupled system
where interactions play an important role in determining the motions
of both the primary and secondary objects.  The bungee jumper is
modeled with a rigid body hierarchy and the bungee cord is modeled
with a spring and mass system.  Because the cord does not
significantly affect the motion of the jumper until after he has left
the platform, we tune the leaping control system with the cord
simulation disabled.  When we are satisfied with the motion for the
leap, we use the two-way coupled system to compute the final motion.

\newcommand{\figureBungee}{
  \centerline{\includegraphics[width=\columnwidth]{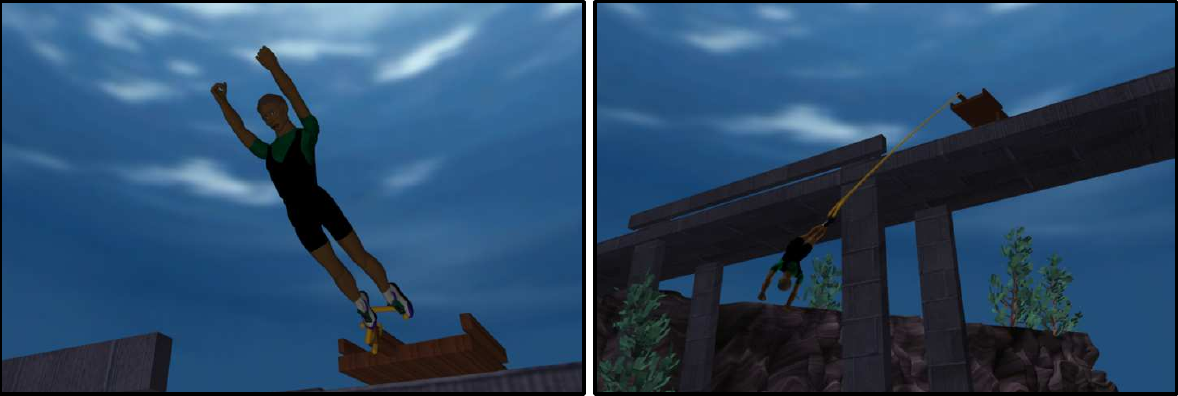}}
  \caption{
    \textbf{Jumper on elastic bungee cord.} 
    The actor's control system causes him to leap from the bridge and
    his fall is arrested by the action of the bungee cord attached to
    his feet.
  }\label{figureBungee}
}
\figureSingle{\figureBungee}


\exsys{Gymnast and Trampoline}
The simulation of a gymnast on a trampoline, shown in
Figure~\ref{figureTramp}, is the most complex of our two-way coupled
examples.  To model this system correctly requires a physically
realistic model of the gymnast, the trampoline, and the interactions
between them, as well as a control system capable of dynamically
balancing the gymnast on the deformable trampoline.  The trampoline is
a spring and mass system.  Parameters for the frame springs and for
the bed of the trampoline were selected to produce deformations
matching those observed in still images and video footage under
similar load conditions\,\cite{Phelps:1990:TSG}.  The control system
is similar to those described previously, but simulated annealing
search techniques were used to automatically determine parameters that
would allow the gymnast to bounce repeatedly.  We chose this approach
because the two-way coupled simulation of the gymnast and the
trampoline was too slow for interactive hand tuning.

\newcommand{\figureTramp}{
  \centerline{\includegraphics[width=\columnwidth]{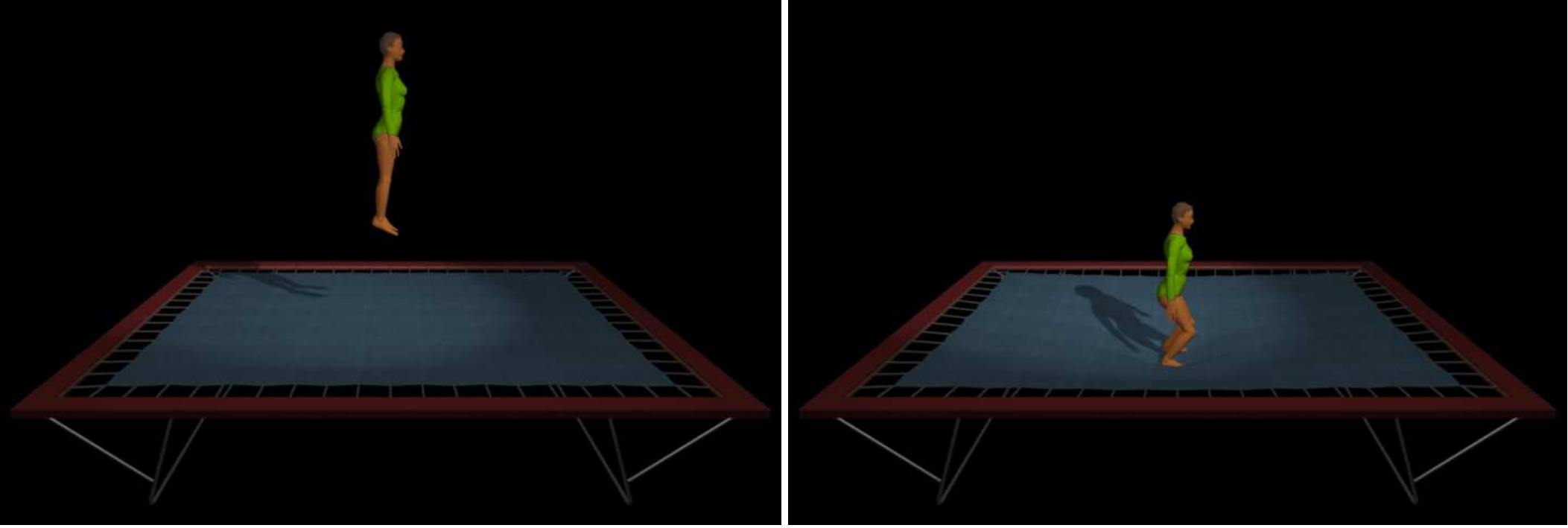}}
  \caption{
    \textbf{Gymnast on a deformable trampoline.}  This system must be
    two-way coupled because the interaction has a significant effect
    on the motions of both systems.  The first image shows the gymnast
    in a layout position prior to landing, the second image shows her
    as she hits the bed of the trampoline.
  }\label{figureTramp}
}
\figureSingle{\figureTramp}


\exsys{Other Examples}
In addition to the examples described above, our coupling methodology
has been used to generate secondary motion for the animated short,
\textit{Alien Occurrence}.  Based on the classic short story
\textit{An Occurrence at Owl Creek Bridge} by Ambrose Bierce, this
animation portrays the sentencing, imagined escape, and final
execution of the main character.  The images in
Figure~\ref{figureOobSamples} show some scenes from the animation with
secondary motion generated using the techniques described in this
paper.

\newcommand{\figureOobSamples}{
  \centerline{\includegraphics[width=\columnwidth]{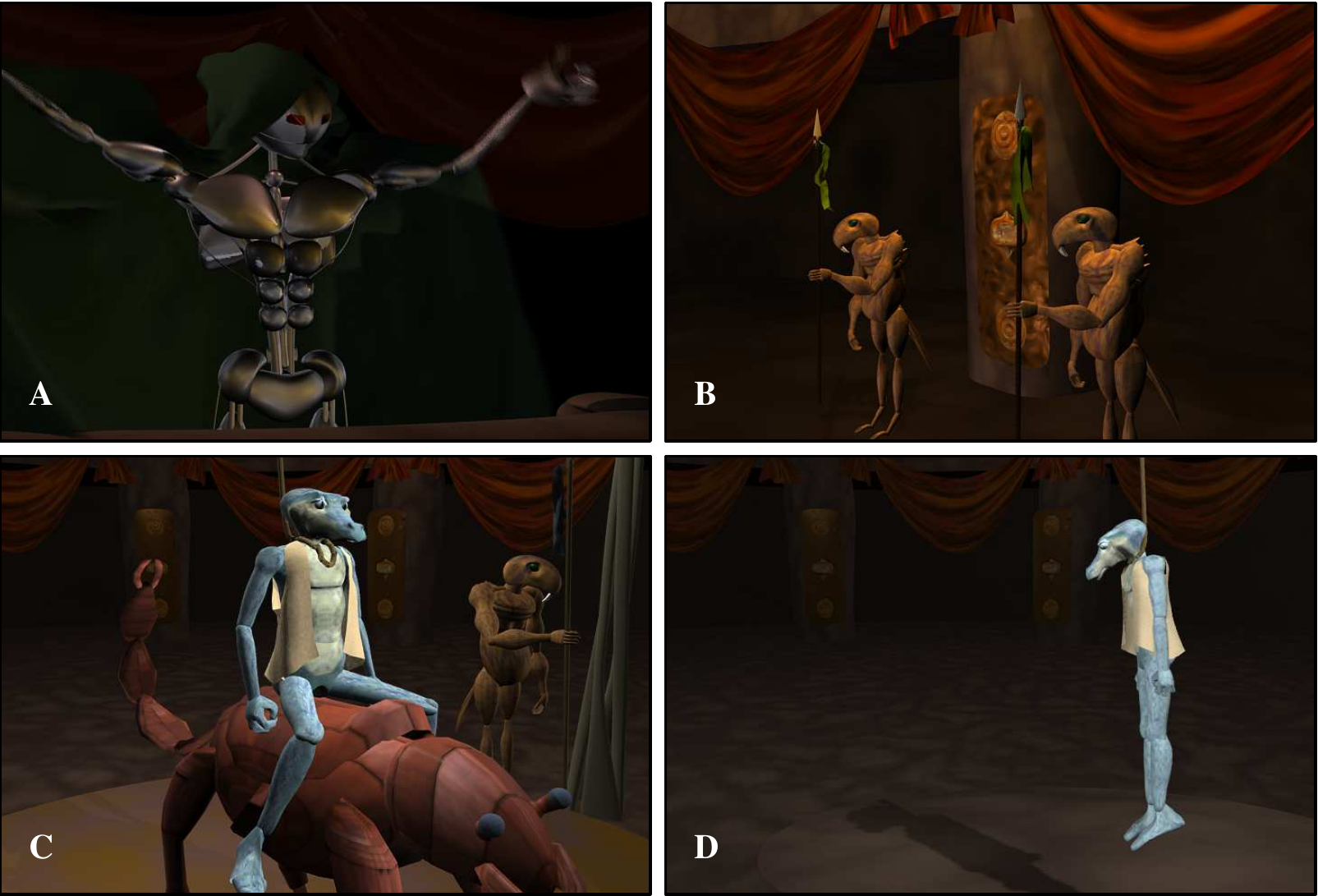}}
  \caption{
    \textbf{Scenes from the animated short \textit{Alien Occurrence}.}
    Secondary elements include: Robe being cast off (A), moving drapes
    in background (A,B,C), tassels on spears (B), vest on condemned
    alien (C,D), and noose (C,D).
  }\label{figureOobSamples}
}
\figureSingle{\figureOobSamples}


\section{Selecting a Coupling Method}

As the preceding examples demonstrate, the best coupling technique
depends on the characteristics of the specific systems and the nature
of the desired effect.  For example, the splash created with a one-way
coupling between the runner's foot and the water is visually
appealing, but if the animator needed to have the runner slip in the
water, a two-way or hybrid coupling would be required.  The decision
process can be facilitated by systematically examining issues such as
complexity, computational speed, interactivity, and stability.  A
decision tree based on an analysis of these factors is shown in
Figure~\ref{figureDecisionTree}.

\newcommand{\figureDecisionTree}{
  \centerline{\includegraphics[width=\columnwidth]{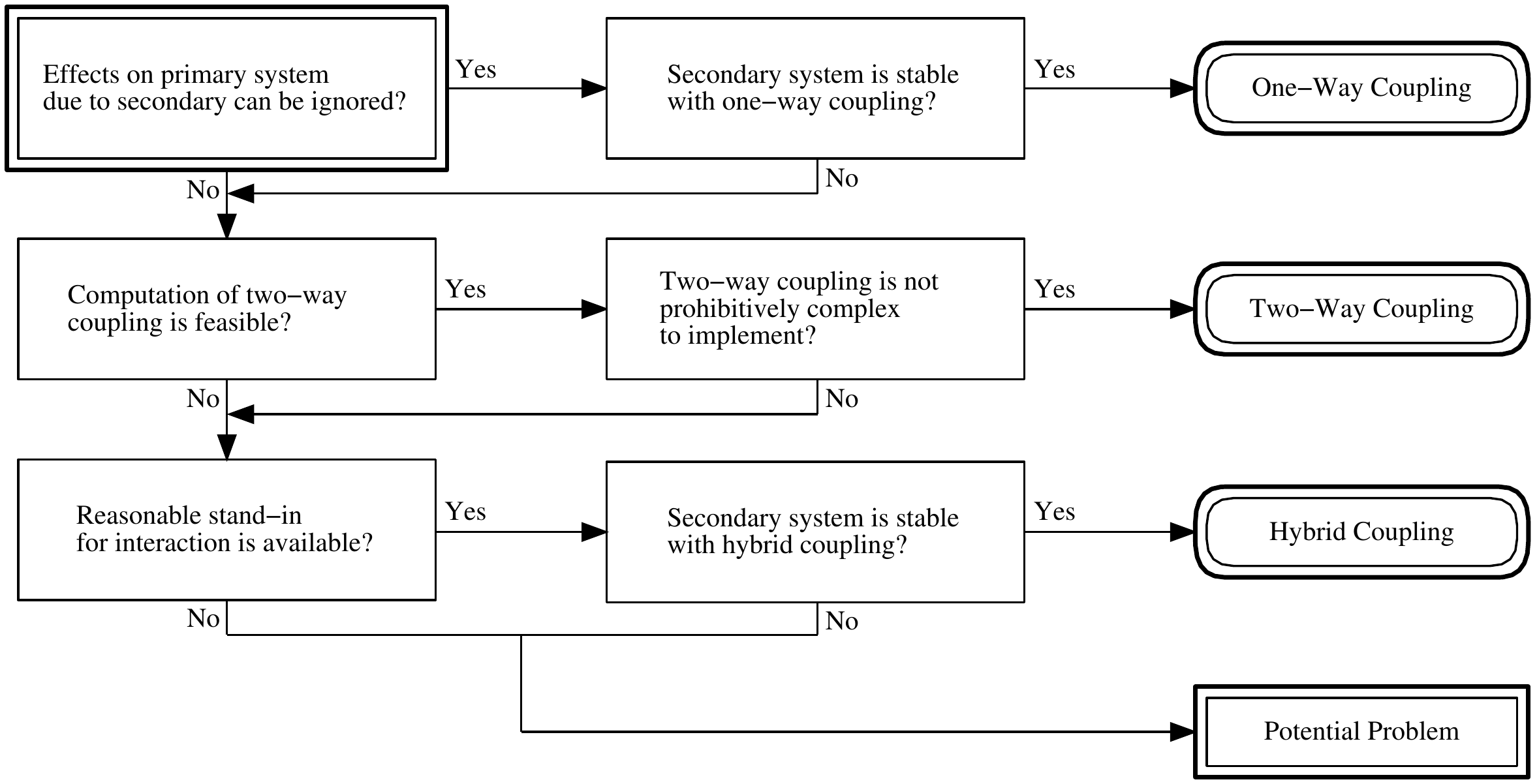}}
  \caption{
    \textbf{Decision tree for coupling selection.}  This diagram
    outlines the process of selecting an appropriate coupling method.
  }\label{figureDecisionTree}
}
\figureSingle{\figureDecisionTree}

\newcommand{\tableForces}{
  \centerline{
    \begin{tabular}{|l|r||l|r||r|r|r||r|r|r|}
      \hline 
      \hline 
      \multicolumn{2}{|c||}{Primary}&\multicolumn{2}{|c||}{Secondary}&\multicolumn{3}{c||}{Force (\frc)}&\multicolumn{3}{c|}{Acceleration (\acc)}\\
      Object    & Mass (\mss) & Object & Mass (\mss) & \multicolumn{1}{c|}{Min} & \multicolumn{1}{c|}{Max} & \multicolumn{1}{c||}{Mean} & \multicolumn{1}{c|}{Min} & \multicolumn{1}{c|}{Max} & \multicolumn{1}{c|}{Mean} \\ 
      \hline
      \hline 
      Ball	&  0.68  & Net        &  0.03  & \spU    0.0  &    59.9  &    15.7  &     0.00 &    88.08 &    23.14 \\ 
      Gymnast   & 64.38  & Trampoline & 20.00  &        60.4  &  5298.8  &  2215.2  &     0.93 &    82.30 &    34.41 \\ 
      Alien	& 46.56  & Vest       &  0.50  &         2.1  &    31.9  &     6.9  &     0.05 &     0.69 &     0.15 \\ 
      Alien	& 46.56  & Noose      &  3.50  & \spD  137.9  &  4055.3  &   575.0  &     2.96 &    87.10 &    12.35 \\ 
      \hline
      \hline
    \end{tabular}
  }
  \caption{
    \textbf{Force and acceleration data from selected simulations.}
    This table shows the interaction forces that occur between two-way
    coupled simulations.  The minimum, maximum, and mean forces are
    computed over the period of time that the objects are in contact or  
    a $0.5$\,second interval in the case of sustained contact.  The 
    accelerations are the effective acceleration on the primary system 
    due to the action of the secondary system.  The rows of this table 
    correspond to Figures~\ref{figureSimBalls}(top
    row),~\ref{figureTramp},~\ref{figureOobSamples}.c,
    and~\ref{figureOobSamples}.d.
  }\label{tableForces}
}
\tableSingle{\tableForces}

If the interaction does not have a significant effect on the primary
system, we can take advantage of the simplicity and speed of one-way
coupling.  An interaction may be insignificant because the primary
object is not influenced by the interaction or because the effect is
contextually unimportant.  The influence on the primary system can be
determined by measuring the effective acceleration due to the sum of
the interaction forces.  Interactions that cause very small
accelerations or accelerations that are overwhelmed by other forces
can probably be ignored.  Table~\ref{tableForces} shows force and
acceleration values for some of the examples presented in this paper.
For the one-way coupled clothing, the acceleration on the primary
system is very small (less than $1$\,m$/$s$^2$), whereas for the
two-way coupled trampoline, the accelerations are much larger
(averaging $34$\,m$/$s$^2$).  The qualitative judgment about whether
an effect is contextually significant is often determined by the
desired level of realism.  Objects that will be part of a busy
background, far away from the camera, or partially obscured do not
require the same level of realism as do objects that are the focus of
attention.

When the interaction is contextually unimportant, system stability may
still rule out the use of one-way coupling.  Because the primary
object's motion is not altered by the interaction, the secondary
system can be driven into unstable configurations or deformed in a
visually unappealing fashion.

When one-way coupling is not feasible, the choice between two-way and
hybrid coupling can be made based on the computational expense and the
complexity of the implementation.  Two-way coupling will result in a
combined system that is, at best, as fast to compute as the slowest
component and possibly much slower because the combined system will
inherit the requirements of both systems.  The greater computational
cost may make the system unusable by increasing the debug cycle time
beyond the user's interactivity threshold.  Two-way coupling may also
be prohibitively complex to implement because of the detailed physical
laws that must be included to model the interaction accurately.

Hybrid coupling is a reasonable choice when a stand-in that cheaply
models the salient elements of the interaction is available.  For
example, our hybrid systems often include vector fields that apply
forces based on the object's position, orientation, and velocity.
Like one-way coupling, hybrid coupling can lead to stability problems,
although adjusting the parameters of the stand-in may alleviate the
problem.

Finally, there are some systems for which the trade-off between
realism and complexity does not yield a reasonable compromise.  For
these systems, one-way coupling is inadequate, two-way coupling is too
expensive, and no suitable stand-in can be devised for hybrid
coupling.  The gymnast and trampoline falls into this category and it
was necessary to employ automated search techniques when developing
the gymnast's controller.

The parameters that determine the appropriate type of coupling may
change during the development cycle. In particular, building two
simulations and the interaction between them in stages allows
programming errors and stability problems to be eliminated before the
full system is assembled.  Furthermore, debugging an active system
with a fast, hybrid coupled system and then switching to two-way
coupling may make designing an effective control system much easier.


\section{Discussion and Conclusions}

In the physical world, all pairs of interacting objects are two-way
coupled and the resulting movement includes a remarkable amount of
perceptible detail.  However, simulation is computationally expensive
and completely simulating even a simple real world scene would be
difficult on current computing hardware.  For this reason, we have
explored three methods of coupling that allow a tradeoff between speed
and realism.  By explicitly considering the interface between
simulations, we have given the animator the ability to choose a
suitable compromise.  This decision about the appropriate level of
coupling is similar to the modeling decision about the level of detail
required for a physical simulation.

While we have focused on the interactions between active and passive
systems, these techniques should be applicable to situations where
both systems are passive or both are active.  The components of the
kites and the initial example of the ball and net demonstrate
passive-to-passive coupling, but we have not shown a system where two
active systems are coupled together, such as would be required for
pairs figure skating.  The simulation of an active-to-active
interaction would be similar to the active-to-passive examples, but
both control systems would have to be robust enough to allow for the
disturbances caused by the changes to the dynamic
systems. Furthermore, when two active simulations are cooperating to
perform a single task, such as a ballet lift, the two control systems
must coordinate the timing and purpose of their actions.

The examples described above demonstrate that our approach of using
coupled simulations is general, can be applied to a wide range of
phenomena, and can add visual richness to an animated scene.  While we
have simulated the secondary motion of many of the objects in the
scene, a number of objects remain motionless.  In some cases, modeling
a few of the moving and flexible objects appears to emphasize the lack
of motion in the others.  Like the progression in models from
wireframe to polygonal to subdivision surfaces, this increase in
fidelity may also increase the viewer's expectations.


\begin{center}
~\\
\noindent
Animations corresponding to the figures in this paper can be in the ancillary files.
\end{center}

\section*{Acknowledgments} 

The authors would like to thank Wayne Wooten for his vaulting and
diving simulations, and Nancy Pollard for her help with the automated
tuning of the trampolinist's control system.  The production of the
animated short, \textit{Alien Occurrence}, involved the dedicated
efforts of many students at the Georgia Institute of Technology's
Graphics, Visualization, and Usability Center.

This project was supported in part by NSF NYI Grant No. IRI-9457621,
Mitsubishi Electric Research Laboratory, and a Packard Fellowship. The
first author was supported by a Fellowship from the Intel Foundation.



\vfill\eject
\end{document}